\documentclass{aa501}
\usepackage{graphicx}
\begin{document}

\def\bril {mJy~beam$^\mathrm{-1}$}
\def\simlt{\mathrel{\rlap{\lower 3pt\hbox{$\sim$}}
        \raise 2.0pt\hbox{$<$}}}
\def\simgt{\mathrel{\rlap{\lower 3pt\hbox{$\sim$}}
        \raise 2.0pt\hbox{$>$}}}

\title{Radio--Optically Selected Clusters of Galaxies}
\subtitle{II. The Cluster Sample
\thanks{Based on observations collected at the European Southern Observatory, 
Chile.}}

\author{A.~Zanichelli \inst{1} 
        \and  R.~Scaramella \inst{2}
        \and G.~Vettolani \inst{1}
        \and  M.~Vigotti \inst{1}
        \and S.~Bardelli \inst{3}
        \and G.~Zamorani \inst{3}
       }
\offprints{A. Zanichelli, \email{azanichelli@ira.bo.cnr.it}}

\institute{Istituto di Radioastronomia -- CNR, Via Gobetti 101, I--40129 
                Bologna, Italy
           \and Osservatorio Astronomico di Roma, via Osservatorio 2, I--00040 
                Monteporzio Catone (RM), Italy
           \and Osservatorio Astronomico di Bologna, via Ranzani 1, I--40127 
                Bologna, Italy
          }

\date{Received ... / Accepted ...}

\abstract{
We present a sample of $171$ candidate groups and clusters of galaxies at 
intermediate redshift over an area of $\approx 550$~sq.~degrees at the South
Galactic Pole, selected by using optically identified radio sources from the 
NRAO VLA Sky Survey as tracers of dense environments.
Out of these $171$ candidates, $76$ have a counterpart in the literature while 
$95$ of them are previously unknown clusters.
This paper presents the cluster selection technique, based on the search of
excesses in the optical surface density of galaxies near identified 
radiogalaxies, and the first spectroscopic results aimed to confirm the 
presence of a cluster. 
Spectroscopy for $11$ candidates led to the detection of $9$ clusters at 
redshift in the range $0.13 \div 0.3$, with estimated velocity dispersions 
ranging from values typical of clusters to those of galaxy groups.
These results show that this technique represents a powerful tool for the 
selection of homogeneous samples of intermediate redshift clusters over a wide 
range of richness.
}

\authorrunning{A. Zanichelli et al.}
\titlerunning{Radio--Optically Selected Galaxy Clusters. II}

\maketitle

\section{Introduction}\label{sec:intr}

Groups and clusters of galaxies are the largest gravitationally bound,
observable structures, and much can be understood about the global 
cosmological properties of the universe by studying their properties -- such as
their dynamical status and evolution, their morphological content and 
interactions with the environment.
To this aim, it is of fundamental importance to gather cluster samples 
representative of different dynamical structures -- from groups to rich 
clusters -- in a wide range of redshift and covering large areas of the sky.

Existing wide-area cluster samples based on visual inspection of optical plates
(Abell et al. \cite{Abell}) or obtained through objective algorithms (EDCC, 
Lumsden et al. \cite{Lumsden}; APM, Dalton et al. \cite{Dalton94}) are limited 
to redshift less than $0.2$, and suffer from the possibility of 
misclassification due to projection effects along the line of sight.
Even more difficult is the detection of groups of galaxies, due to their low 
density contrast with respect to field galaxy distribution.
In the optical band, cluster samples at higher $z$ have been built over 
selected areas of few square degrees (Postman et al. \cite{Postman}; Scodeggio 
et al. \cite{Scodeggio}).
Alternatively, the X-ray emission of the hot intracluster medium has been 
widely used to build distant cluster samples, but this technique suffers from 
the limited sensitivity of wide-area X-ray surveys and from the possibility of 
evolutionary effects (Gioia et al. \cite{Gioia}; Henry et al. \cite{Henry}; 
RDCS, Rosati et al. \cite{Rosati}).

A different approach -- complementary to purely optical or X-ray cluster
selection methods -- is the use of radiogalaxies as suitable tracers of dense 
environments.
Faranoff--Riley I and II radio sources have been shown to inhabit different 
environments at different epochs  and proved to be efficient tracers of galaxy 
groups and clusters (Prestage \& Peacock \cite{Prestage}; Hill \& Lilly 
\cite{Hill}; Allington--Smith et al. \cite{Allington--Smith}; Zirbel 
\cite{Zirbel97}).
FRI sources are found on average in rich groups or clusters at any redshift,
and are associated with elliptical galaxies, with the most powerful FRI often 
hosted by a cD or double nucleus galaxy.
FRII radio sources are typically associated with disturbed ellipticals
and they avoid rich clusters at low $z$ (Zirbel \cite{Zirbel96}).

Since there is no significant correlation between the radio properties of 
galaxies within a cluster with the cluster X-ray luminosity (Feigelson et al.
\cite{Feigelson}; Burns et al. \cite{Burns}), or richness (Zhao et al. 
\cite{Zhao}; Ledlow \& Owen \cite{Ledlow}), as well as between the properties 
of group members and the radio characteristics of the radiogalaxies (Zirbel 
\cite{Zirbel97}), radio selection should not impact on the X-ray or optical 
properties of the clusters found in this way.

Radiogalaxies can thus be used to study the global properties of galaxy groups 
and clusters, such as their morphological content, dynamical status and number 
density, as well as the effect of the environment on the radio emission
phenomena.

We used the NRAO VLA Sky Survey (NVSS, Condon et al. \cite{Condon}) publicly 
available data to build a sample of radio-optically selected clusters through 
optical identifications of radio sources and search of excesses in the surface 
density of galaxies around these radiogalaxies.
The NVSS survey offers indeed an unprecedented possibility to study a 
wide-area, homogeneous sample of radio sources down to relatively low flux 
levels, together with a positional accuracy suitable for optical 
identifications.

In a previous paper (Zanichelli et al. \cite{Zanichelli}, hereafter Paper I) 
we described how we extracted a radio source catalogue  from the NVSS maps and 
the optical identification procedure that led to the compilation of a 
radiogalaxy sample.
In this paper we discuss the cluster finding method used for the compilation
of a new sample of candidate groups and clusters of galaxies, and present the
first observational results that spectroscopically confirmed the presence of a 
group or cluster for $9$ out of the $11$ successfully observed candidates.

This paper is structured as follows: in Sect.~\ref{sec:data} we give a summary 
of the properties of the radio and optical data samples used for the cluster 
search. In Sect.~\ref{sec:clussel} we describe the cluster finding method. The 
new sample of candidate clusters is presented in Sect.~\ref{sec:clussample}.
Spectroscopic observations of a subsample of candidate clusters, aimed to 
obtain an observational confirmation of the presence of a cluster are 
presented and discussed in Sects.~\ref{sec:observations} 
and ~\ref{sec:results}.

\section{The data}\label{sec:data}

In the following two Sects. we recall the properties of the radiogalaxy sample 
and of the optical galaxy catalogue that have been used during this search. For
more details on the radio source catalogue and the definition of the 
radiogalaxy sample, refer to Paper I.

\subsection{The radiogalaxy sample}\label{sec:rgsample}

The radio source catalogue has been extracted from $31$ maps of the $1.4$~GHz
NRAO VLA Sky Survey (Condon et al. \cite{Condon}) and consists of $13\,340$ 
pointlike and $2662$ double radio sources down to a flux limit of $2.5$~\bril~ 
over an area of $\approx 550$~sq.~degrees at the South Galactic Pole.

Optical identifications of NVSS radio sources have been made with galaxies
brighter than $b_\mathrm{J} = 20.0$ in the EDSGC catalogue (Nichol et al. 
\cite{Nichol}) using a search radius of $15\arcsec$, i.e. $\approx 3\sigma$ 
positional accuracy for the faintest sources.
The initial sample of optical counterparts consists of $1288$ radiogalaxies, 
$926$ of them having a pointlike radio morphology at the NVSS resolution of 
$45\arcsec$.

As shown in Table 2 of Paper I, the contamination level due to spurious
identifications varies according to the radio morphological classification,
ranging from about $16\%$ for the lists of optical counterparts of pointlike 
radio sources and ``close'' radio pairs (separation between components 
$D \le 50\arcsec$), to about $28\%$ for the list of optical counterparts of 
``wide'' radio doubles ($50\arcsec < D < 100\arcsec$).

In order to obtain a more reliable sample, the radiogalaxy data set used in the
search of candidate clusters has been selected among these optical 
identifications on the basis of radio-optical distance and galaxy magnitude.
The  uncertainty in the optical identification sample is indeed the only source
of contamination that can be limited when selecting cluster candidates by 
looking for excesses in surface galaxy density near the identified 
radiogalaxies.
Other contamination terms -- like the probability of detecting a candidate by 
chance coincidence of the radiogalaxy position with an optical density excess, 
or the possibility that the optical excess itself is intrinsically spurious, 
i.e. due to chance superpositions of galaxies along the line of sight -- 
cannot in fact be reduced unless one knows the redshift distribution of the 
galaxies.

From the initial sample of optical counterparts we thus selected those 
radiogalaxies having $d_\mathrm{r-o} \le 7\arcsec$. This constraint introduces 
a selection effect against faint sources in the radio sample, whose positional 
uncertainty is typically $\sim 5\arcsec$.

Furthermore, as our aim is to select candidate clusters at intermediate
redshifts, we discarded those radiogalaxies brighter than magnitude
$b_\mathrm{J} = 17.5$.
In fact, considering the magnitude -- redshift relation typical of 
radiogalaxies obtained in the R band  by Grueff \& Vigotti (\cite{Grueff}), and
using color indexes for elliptical galaxies given in Frei \& Gunn 
(\cite{Frei}), this cut in $b_\mathrm{J}$ magnitude corresponds to a redshift 
lower limit of $z \simgt 0.1$.

With these constraints, the final radiogalaxy sample that has been taken into 
account for the search of candidate clusters consists of $661$ radiogalaxies, 
and the mean, expected contamination level due to spurious optical 
identifications has been lowered to about $10\%$.

\subsection{The galaxy catalogue}\label{sec:galcatalog}

The Edinburgh--Durham Southern Galaxy Catalogue (EDSGC, Nichol et al. 
\cite{Nichol}) lists $\approx 1.5 \times 10^6$ galaxies over a contiguous area 
of $\sim 1200$~sq.~degrees at the South Galactic Pole.
About one half of this area is currently covered by our radiogalaxy sample and 
has been considered for the search of cluster candidates.

The EDSGC has been obtained from COSMOS scans of IIIa--J ESO/SERC plates at 
high galactic latitude ($\mid b_\mathrm{II}\mid \ge 20\degr$). The automated 
star-galaxy separation algorithm used for the EDSGC guarantees a completeness 
$> 95\%$ and a stellar contamination $< 12\%$ down to magnitudes 
$b_\mathrm{J}= 20.0$. 

Magnitudes have been calibrated via CCD sequences, providing a plate-to-plate
accuracy of $\Delta b_\mathrm{J}\simeq 0.1$ and an rms plate zero-point offset 
of $0.05$ magnitudes. 

The EDSGC incompleteness starts to exceed the $5\%$ only above 
$b_\mathrm{J} =20.5$ (Collins et al. \cite{Collins92}). When looking for 
candidate clusters we thus decided to make optical galaxy counts down to the 
magnitude limit $b_\mathrm{J}=20.5$: as the radiogalaxy sample reaches 
$b_\mathrm{J}=20.0$, this choice makes it possible to point out also those 
regions of high galaxy surface density associated to the optically faintest 
radio sources in our sample.

\section{Joint radio--optical cluster selection}\label{sec:clussel}

The cluster finding method we adopted is based on optical counts of galaxies in
cells, followed by a smoothing of these counts with a Gaussian function and by
the definition of a detection threshold for the selection of significative 
excesses in the surface galaxy density.
The density peaks that are found near an optically identified NVSS radio source
are included in the cluster sample.

Both to keep in evidence any possible inhomogeneities in optical counts and to 
make the data handling easier, we divided the EDSGC galaxies brighter than 
$b_\mathrm{J} = 20.5$ in $21$ adjacent sky maps corresponding to the 
$5\degr \times 5\degr$ central regions of the UKST plates that cover the radio 
source catalogue area.

The detection threshold is built in terms of the mode and rms of galaxy 
counts over each of these sky regions. Despite the possibility of small 
intra-plate variations in the photometric accuracy of the optical data, we 
considered the choice of a ``local'' threshold for each plate preferable to a 
``global'' one, over the whole sky region, as the latter choice would introduce
in the cluster sample incompleteness effects that depend on the cluster 
location in the sky.

The optical count matrix for each $5\degr \times 5\degr$ sky region has been 
built by defining a regular grid of $600 \times 600$ cells and by counting 
galaxies in these cells.
The size of each cell has been chosen to be $30\arcsec \times 30\arcsec$, to 
optimize the statistics of galaxy counts as well as to point out the presence 
of structure in the spatial distribution of galaxies.

Since radiogalaxies tend to reside in different environments -- from groups of 
galaxies to rich clusters -- depending on their Faranoff--Riley morphological 
classification (Zirbel \cite{Zirbel97}), a careful choice of the size for the 
smoothing function is needed to avoid selection effects in favour of a 
particular environment.
Too large a size could translate into a lack of detections of distant clusters,
whose angular sizes are small. A small size could resolve a nearby cluster in 
many substructures thus leading to the spurious detection of many candidates
relative to the actual cluster, or, if the optical surface density 
excess in each substructure is less than the selected threshold (see below), 
could lead to a lack of detections. This last case is the most likely for 
clusters at moderate $z$ having irregular morphologies, like the Abell I types
(Abell \cite{Abell58}), where subclumps in the galaxy distribution are seen.
Given the redshift range we expect to cover with our cluster sample, we decided
to adopt, for the smoothing of the optical counts, a circular Gaussian function
of FWHM$ = 2\arcmin$, which is about half an Abell radius at $z=0.4$.

\begin{figure*}
\vspace{17truecm}
\caption{
Smoothed matrix of the optical counts corresponding to the
$5\degr \times 5\degr$ sky region of the UKST plate 412. Different grey levels 
correspond to regions of high galaxy density; superimposed contours are given 
as $2,3,...\sigma_\mathrm{gal}$ above $m_\mathrm{gal}$. 
On this plate there are $13$ candidate clusters matching the selection 
criteria described in Sect.~\ref{sec:clussel}, marked with solid circles.
The $44$ NVSS radio sources optically identified in this sky region are marked 
with diamonds, except for the $2$ residing in candidates associated with 
clusters known from the literature (A2878 and E536), that are marked with 
asterisks.
Big dashed circles, with radius equal to one Abell radius, show ACO/Abell 
clusters on this plate, while the dotted circle marks the cluster EDCC 536.}
\label{fig1}
\end{figure*}

We then looked for significative excesses in the optical surface galaxy 
density: for each smoothed matrix, we determined the mode $m_\mathrm{gal}$ and 
rms $\sigma_\mathrm{gal}$ of the optical surface density.

The threshold we adopted for the detection of density excesses is defined on 
each smoothed plate in terms of $m_\mathrm{gal}$ and $\sigma_\mathrm{gal}$ as
$n_\mathrm{threshold} = m_\mathrm{gal} + 3\sigma_\mathrm{gal}$, that is we 
consider significative only those peaks where the galaxy surface density 
exceeds by at least $3\sigma_\mathrm{gal}$ the value of the mode determined
over the whole plate.
This choice can introduce selection effects in favour of ``core-dominated'', 
regular clusters, and against irregular ones, where galaxies are less 
concentrated in the cluster core itself.

Finally, from this list of significant peaks we selected only those for which a
radiogalaxy belonging to the considered smoothed matrix is found at a maximum 
distance of $4\arcmin$ from the density peak position.
To determine the list of density peaks, no constraint has been set on the 
number of connected cells above the threshold. Radiogalaxies themselves are not
required to belong to pixels whose optical surface density is above 
$n_\mathrm{threshold}$.
Given the definition of the Abell radius, 
$R_\mathrm{A} =  {1.7\arcmin \over z}$, this search distance corresponds to the
Abell radius of a cluster at $z \sim 0.45$.

In the case of nearer clusters, this choice will favour the selection of those 
candidates where the radiogalaxy is located in the central region of the
cluster.
A larger value of the search distance would however increase the probability of
detecting spurious associations between density peaks and radiogalaxies.

In Fig.~\ref{fig1} we show as an example the smoothed matrix relative to 
the UKST plate 412: regions with higher surface galaxy density are represented 
by increasing grey levels; the superimposed contours are given as 
$2,3,...\times \sigma_\mathrm{gal}$ above $m_\mathrm{gal}$.
The $44$ radiogalaxies having $17.5 \le b_\mathrm{J} \le 20.0$ and
$d_\mathrm{r-o} \le 7\arcsec$ present in this sky region are shown as well, 
marked with diamonds, except for the two associated with a known candidate 
cluster (see Sect.~\ref{sec:clussample}), that are marked with an asterisk.
The $13$ radio--optically selected cluster candidates are marked as small
circles around the position of the associated radiogalaxy.
For clarity, only the ACO/Abell and EDCC clusters found in this sky region
are plotted in Fig.~\ref{fig1}: they are marked with big circles, whose 
radius is equal to the cluster Abell radius.
The fact that the clusters A2878, 2904 and E536 do not seem very conspicuous in
this galaxy density map can be explained in terms of the different optical data
set and the different scale used by Abell et al. (\cite{Abell}) and Lumsden et 
al. (\cite{Lumsden}) to look for overdensities in the galaxy distribution.

\begin{figure*}
\centering
\includegraphics[width=17cm]{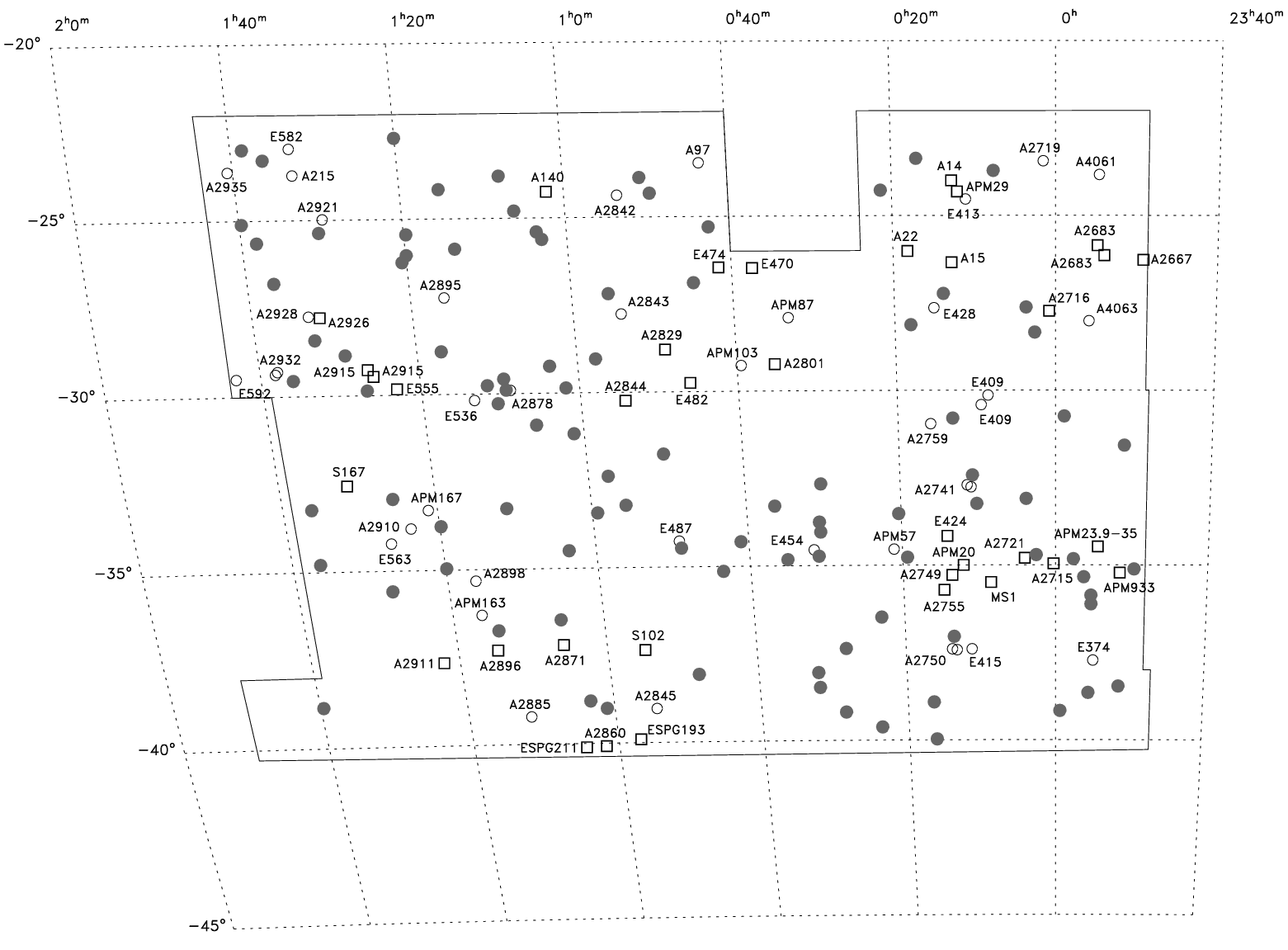}
\caption{
Sky distribution of the candidate clusters in our sample over the region
defined by the NVSS catalogue ($\approx 550$~sq.~degrees). Empty symbols refer 
to $76$ candidates associated to clusters known in the literature, for which 
the name is marked; circles refer to known clusters with estimated redshift, 
while squares mark known clusters with measured redshift. Filled symbols refer 
to the $95$ previously unknown candidates.}
\label{fig2}
\end{figure*}

\section{The candidate cluster sample}\label{sec:clussample}

By applying the cluster finding method described in the previous Section, we 
obtained a sample of $171$ candidate clusters associated to NVSS radio sources
identified with $d_\mathrm{r-o} \le 7\arcsec$ with EDSGC galaxies of magnitude 
$17.5 \le b_\mathrm{J} \le 20.0$.
Among these candidates, $123$ are associated to NVSS pointlike sources, $23$ to
NVSS ``close'' double sources, and $25$ to NVSS ``wide'' doubles. The sample 
covers an area of $\approx 550$~sq.~degrees at the South Galactic Pole and the 
uncertainty on the candidate cluster position is $30\arcsec$.

As a further step in the compilation of the cluster sample we looked for
candidates in common with other cluster catalogues, considering a candidate as 
already ``known'' if the radiogalaxy and the optical centroid are found inside 
an Abell radius from the cluster centre.

Out of $171$ candidates, $76$ were found to be associated with a known cluster 
according to the above definition.
In six cases two candidates are associated with the same known cluster.
We detected $2$ Abell poor clusters with measured redshift, $40$ out of the 
$128$ ACO/Abell (Abell et al. \cite{Abell}) rich clusters present in this sky 
region ($31$ of them being listed in the EDCC as well), $16$ EDCC (Lumsden et 
al. \cite{Lumsden}) clusters, $9$ clusters from the APM catalogue (Dalton et 
al. \cite{Dalton94}, \cite{Dalton97}), $2$ groups selected from the ESO Slice 
Project survey (Ramella et al. \cite{Ramella}). Finally, in one case 
the candidate corresponds to a cluster identified with an X-ray source in the 
Einstein Medium Sensitivity Survey (Stocke et al. \cite{Stocke}).

To evaluate $R_\mathrm{A}$, we used the measured cluster redshift when 
available; otherwise, we used the estimated redshift or, in the case of EDCC 
clusters, values of $R_\mathrm{A}$ estimated by the catalogue authors.
In thirty-six cases the association with previously known clusters has been
made on the basis of a measured redshift in the literature, while for the 
other $40$ only an estimated redshift is available.

The detection of these known clusters can be interpreted as a further
indication that this radio-optical selection method is powerful in the search 
of cluster candidates.

Among the $76$ known clusters, $57$ host a pointlike NVSS radio source, while 
$11$ and $8$ are respectively associated to ``close'' and ``wide'' double radio
sources.

In Fig.~\ref{fig2} the sky distribution of the known candidates (empty 
symbols marked with the cluster name) and of those candidates without a 
counterpart in the literature (filled symbols) is shown.

The use of this bivariate radio--optical selection method, based on the
condition that an optical excess is considered as a reliable candidate cluster 
only if it is associated with a radiogalaxy, makes it possible to detect 
cluster candidates whose reliability, in terms of their optical surface density
alone, would be in many cases too low to be included in catalogs based on pure 
optical selection methods.

The total contamination present in the cluster sample, due to the probability 
of chance coincidence between a radiogalaxy and an excess in the optical 
surface galaxy density, has been estimated as follows: we repeated the search 
of cluster candidates coupling the smoothed matrix relative to each UKST plate 
with the radiogalaxies belonging to another plate.
By applying the same cluster finding criteria described above, we found a 
contamination percentage of $28\%$ in our sample of candidate clusters at 
intermediate redshifts.
As the actual radiogalaxy sample instead of a random-generated one has been 
used, this is an estimate of the effective contamination, that is also the 
contamination term due to the presence of spurious radio-optical associations
among the radiogalaxies is taken into account.

An assessment of the reliability of the method is provided by the association 
with known clusters and measurements on new candidates. Therefore, in the next 
two Sects. we briefly report on those candidates hosting a radiogalaxy whose 
redshift is known from previous surveys.
This search has been made using the NASA Extragalactic Database 
\footnote{The NASA/IPAC Extragalactic Database (NED) is operated by the Jet 
Propulsion Laboratory, California Institute of Technology, under contract with 
the National Aeronautics and Space Administration.}.
In Sect.~\ref{sec:observations} and ~\ref{sec:results} the results obtained 
from a spectroscopic observative run for a first set of new cluster candidates 
are presented. The whole candidate cluster sample will be presented in a 
following paper.

\subsection{Previously known clusters}\label{sec:knownclus}

The naming convention for cluster candidates in our sample is as follows: first
digits identify the number of the UKST plate on which the candidate has been 
found; 
letters are used to distinguish among the various radio morphologies (pointlike
or double sources) and to identify candidates associated to more than one 
radiogalaxy. Last digits in the name are the sequential number of the radio 
source on that UKST plate.

\noindent
{\bf 295BN07}: this candidate is found to correspond to A2860, whose measured 
redshifts is $z=0.105800$ (Struble \& Rood, \cite{Struble}). The radiogalaxy 
lies at $z = 0.10757 \pm 0.00018$ (Vettolani et al. \cite{Vettolani98}) so that
295BN07 is considered actually coincident with a known cluster.

\noindent
{\bf 295D24}: the candidate is found inside one Abell radius from the ESP Group
193 (Ramella et al., \cite{Ramella}). The radiogalaxy has been detected at 
$z = 0.05547 \pm 0.00007$ in the ESP survey (Vettolani et al. 
\cite{Vettolani98}). ESP group 193 lies at redshift $z=0.05444 \pm 0.001$ so 
that the association is considered real.

\noindent
{\bf 352N25}: the candidate has been found to be associated with the ACO 
cluster 
A2871. In A2871 the presence of $2$ galaxy systems at different redshifts has 
been detected (Katgert et al., \cite{Katgert96}) respectively at $<z> = 0.114$ 
($14$ galaxies) and $<z> = 0.122$ ($18$ galaxies). The radiogalaxy has measured
redshift from the ENACS survey $z = 0.11415 \pm 0.0003$ (Katgert et al. 
\cite{Katgert98}) so 352N25 seems to be associated with one of these two 
substructures.

\noindent
{\bf 411N35}: the radiogalaxy in this candidate lies inside an Abell radius 
from the center of the EDCC cluster E482, whose measured redshift is 
$z = 0.108040$. The radiogalaxy has $z = 0.07673 \pm 0.0003$ (Collins et al. 
\cite{Collins95}); 411N35 is thus not actually coincident with E482. However, 
measured velocities in the E482 field include a set of galaxies with 
$cz \sim 22700~{\rm km~s}^\mathrm{-1}$, indicating the possible presence of a 
superimposed structure at about the same redshift as the radio source.

\subsection{New cluster candidates}\label{sec:newclus}

From the publicly available data of the Las Campanas Redshift Survey (Shectman 
et al. \cite{Shectman}) we obtained the redshift of two radiogalaxies 
associated with the new candidate clusters 297BN04 
($cz = 53460 \pm 58~{\rm km~s^{-1}}$) and 293D22 
($cz = 41695 \pm 95~{\rm km~s^{-1}}$).
At $\sim 2\arcmin$ from the radiogalaxy in 297BN04 we found a galaxy with
measured velocity $cz = 53319~{\rm km~s^{-1}}$.
Finally, a further redshift has been found in the ESP survey for the 
radiogalaxy associated to the candidate 294N04: $cz = 43097~{\rm km~s^{-1}}$ 
(Vettolani et al. \cite{Vettolani98}).
No redshift data for other galaxies near these radiogalaxies are available, so 
these velocity measurements cannot be used to confirm or not the presence of 
a cluster. 

\begin{figure*}
\begin{centering}
\begin{tabular}{cc}
\vspace{22.5truecm}
\end{tabular}
\caption{$10$ min, r-Gunn images taken with the $3.6$ m ESO telescope for $12$
cluster candidates. From left to right and top to bottom: 294N15, 295N35,
349N02, 350N71, 352N47, 352N63. Objects for which we acquired the spectra
are marked with horizontal or vertical lines; the radiogalaxy is marked by
an arrow.}
\end{centering}
\end{figure*}
\begin{figure*}
\begin{centering}
\addtocounter{figure}{-1}
\begin{tabular}{cc}
\vspace{22.5truecm}
\end{tabular}
\caption{(Continued) From left to right and top to bottom: 352N75, 409N03,
409N15, 409N44, 412N23, 475N50.}
\label{fig3}
\end{centering}
\end{figure*}

\section{Optical observations}\label{sec:observations}

A first set of $14$ visually good candidate clusters associated with pointlike 
NVSS radio sources have been observed with the $3.6$m ESO telescope at La 
Silla, with the EFOSC1 spectrograph in multislit mode.
Moreover, photometry of each field in the r-Gunn filter has been acquired 
during the first night of observation to achieve magnitude $r \simeq 22.5$ -- 
corresponding to roughly $b_\mathrm{J} \sim 24$ -- with a photometric accuracy 
better than $0.2$ mag.

The targets for spectroscopic observations were chosen on the basis of these 
photometric observations. For $12$ candidate clusters the spectra of the 
radiogalaxy and about $10 - 14$ companions were acquired.
Observations were made with the $B300$ grism, characterized by a wavelength 
range $3740 - 6950~\mathrm{A}$, central wavelength 
$\lambda_\mathrm{c} = 5250~\mathrm{A}$ and
dispersion $230~\mathrm{A~mm^{-1}}$. The slit width was chosen to be 
$2.1\arcsec$ while the slit length varied in order to optimize the number of 
acquired spectra. The resolution on the spectra was about $20~\mathrm{A}$.

Exposures of He-Ar lamps for wavelength calibration have been acquired through
the same masks used in the scientific exposures. Spectroscopic dome flats 
proved not to be useful during the data reduction phase: due to the low quality
of the slit profiles achieved with the mask Punching Machine, the flatfielding
process did not significantly help in the extraction of spectra. As we did not
apply flat field correction, the obtained spectra are not flux calibrated.

\subsection{Data reduction}\label{sec:datared}

Multislit spectroscopic data reduction has been made interactively by means of
the IRAF package. Reduction steps involved bias subtraction, spectra 
extraction, spectra wavelength calibration. We found no need to correct for 
dark current. This procedure has been applied both to astronomical and to 
calibration lamps exposures.

We acquired a total number of $129$ galaxy spectra, plus $4$ stars, for $12$ 
successfully observed candidates, and determined the velocity of galaxies from 
absorption and, in a few cases, emission lines by means of the RVSAO package.
None of the radiogalaxies we observed show emission lines in their spectra.
Templates for the cross-correlation consist of $8$ galaxy and $8$ star spectra
known from previous observative programs: the $8$ galaxies were observed during
the Edinburgh--Milano Cluster Survey (Collins et al. \cite{Collins95}) with 
EFOSC1 and with the same spectral resolution as our observations. The $8$ stars
come from ESO Slice Project (Vettolani et al. \cite{Vettolani97}) observations
with the fiber spectrograph OPTOPUS.

To allow for cleaning of cosmic rays, multiple exposures were taken for each 
field, for a total exposure time varying from $40$ to $60$ min. In 
Fig.~\ref{fig3} the direct imaging exposures of these $12$ candidates are 
shown, together with the targets we selected for spectroscopy.

In Table~\ref{tab:allvel} the r-Gunn magnitudes, the measured velocities and 
their associated errors are given for each galaxy we observed in the selected 
candidate clusters. As can be seen from Table~\ref{tab:allvel}, for $22$ out of
the $129$ observed galaxies the S/N was not good enough to measure the 
redshift.
The typical $r$ magnitude for galaxies with measured redshift is $18.7$. Stars 
are marked with an asterisk. The high errors associated to some velocity 
measurements are mainly due to the low spectral S/N.

\begin{table}
\begin{flushleft}
\caption{Measured velocities and r-Gunn magnitudes for galaxies in the $12$ 
observed cluster candidates. Notes: ``R'' = radiogalaxy; ``E'' = emission line 
galaxy; ``$\ast$'' = star.
\label{tab:allvel} }
\begin{tabular}{l@{\extracolsep{3mm}}l@{\extracolsep{6mm}}r@{\extracolsep{6mm}}c@{\extracolsep{6mm}}c@{\extracolsep{6mm}}c} 
\hline
\hline\noalign{\smallskip}
NAME & N & $m_\mathrm{r}$ & $v$ & $\sigma_\mathrm{v}$~~~ & Notes \\
& & & ${\rm (km~s^{-1})}$ & ${\rm (km~s^{-1})}$ & \\
\hline
\hline\noalign{\smallskip}
{\bf 294N15} & 1 & 19.69 & 89125 & $\pm$ 86 & \\
& 2 & 19.09 & 92061 & $\pm$ 60 & \\
& 3 & 17.67 & 19572 & $\pm$ 300 & E \\
& 4 & 17.42 & 90363 & $\pm$ 41 & R \\
& 5 & 19.78 & 85666 & $\pm$ 122 & \\
& 6 & 20.10 & - & - & \\
& 7 & 18.55 & 89065 & $\pm$ 50 & \\
& 8 & 20.05 & 89540 & $\pm$ 46 & \\
& 9 & 15.62 & 21186 & $\pm$ 48 & \\
& 10 & 20.53 & 61998 & $\pm$ 100 & \\
& 11 & 20.13 & - & - & \\
& 12 & 20.33 & 90960 & $\pm$ 259 & \\
\hline\noalign{\smallskip}
{\bf 295N35} & 1 & 20.90 & - & - & \\
& 2 & 19.50 & 23511 & $\pm$ 130 & \\
& 3 & 19.68 & 31000 & $\pm$ 300 & \\
& 4 & 19.08 & 79401 & $\pm$ 309 & \\
& 5 & 19.53 & 97000 & $\pm$ 300 & \\
& 6 & 20.10 & 78568 & $\pm$ 107 & \\
& 7 & 17.84 & 79110 & $\pm$ 379 & R \\
& 8 & 21.29 & 14800 & - & E \\
& 9 & - & - & - & \\
& 10 & 20.51 & 98925 & $\pm$ 114 & \\
& 11 & 18.38 & 79863 & $\pm$ 198 & \\
\hline\noalign{\smallskip}
{\bf 349N02} & 1 & 19.59 & 34237 & $\pm$ 300 & \\
& 2 & 18.80 & 29093 & $\pm$ 130 & \\
& 3 & 18.81 & 63333 & $\pm$ 85 & \\
& 4 & - & - & - & \\
& 5 & 18.42 & - & - & $\ast$ \\
& 6 & 16.03 & 33698 & $\pm$ 49 & R \\
& 7 & 18.83 & - & - & $\ast$ \\
& 8 & 19.69 & 123966 & $\pm$ 300 & E \\
& 9 & 20.84 & 146819 & $\pm$ 300 & E \\
& 10 & 21.45 & - & - &  \\
\hline\noalign{\smallskip}
{\bf 350N71} & 1 & 16.97 & 42919 & $\pm$ 74 & \\
& 2 & 17.16 & 42586 & $\pm$ 86 & \\
& 3 & 20.62 & 80696 & $\pm$ 112 & \\
& 4 & 17.73 & 69681 & $\pm$ 169 & \\
& 5 & 16.23 & 56290 & $\pm$ 80 & \\
& 6 & 18.79 & 56099 & $\pm$ 280 & \\
& 7 & 17.80 & 69904 & $\pm$ 109 & \\
& 8 & 17.91 & 70484 & $\pm$ 82 & R \\
& 9 & 18.44 & 70824 & $\pm$ 119 & \\
& 10 & 18.36 & 70060 & $\pm$ 121 & \\
\hline\noalign{\smallskip}
{\bf 352N47} & 1 & 19.77 & 51980 & $\pm$ 137 & \\
& 2 & 16.68 & 51988 & $\pm$ 114 & R \\
& 3 & 19.60 & 81249 & $\pm$ 258 & \\
& 4 & 18.06 & 51233 & $\pm$ 150 & \\
& 5 & 20.85 & - & - & \\
& 6 & 18.70 & 52499 & $\pm$ 150 & \\
& 7 & 20.02 & 86464 & $\pm$ 300 & \\
& 8 & 19.63 & 63814 & $\pm$ 220 & \\
& 9 & 19.90 & 97518 & $\pm$ 172 & \\
& 10 & 19.81 & - & - & \\
\hline\noalign{\smallskip}
\end{tabular}
\end{flushleft}
\end{table}
\begin{table}
\begin{flushleft}
\addtocounter{table}{-1}
\caption{(Continued).}
\begin{tabular}{l@{\extracolsep{3mm}}l@{\extracolsep{6mm}}r@{\extracolsep{6mm}}c@{\extracolsep{6mm}}c@{\extracolsep{6mm}}c}
\hline
\hline\noalign{\smallskip}
NAME & N & $m_\mathrm{r}$ & $v$ & $\sigma_\mathrm{v}$~~~ & Notes \\
& & & ${\rm (km~s^{-1})}$ & ${\rm (km~s^{-1})}$ & \\
\hline
\hline\noalign{\smallskip}
{\bf 352N63} & 1 & 18.23 & 54587 & $\pm$ 71 & \\
& 2 & 19.01 & 55881 & $\pm$ 189 & \\
& 3 & 21.28 & - & - & \\
& 4 & 20.07 & - & - & \\
& 5 & 16.73 & 54698 & $\pm$ 118 & \\
& 6 & 18.38 & 54699 & $\pm$ 76 & R \\
& 7 & 18.29 & 53486 & $\pm$ 84 & \\
& 8 & 18.74 & 55275 & $\pm$ 78 & \\
& 9 & 20.60 &- & - & \\
& 10 & 20.32 & - & - & \\
& 11 & 19.85 & - & - & \\
\hline\noalign{\smallskip}
{\bf 352N75} & 1 & 19.57 & - & - & \\
& 2 & 18.91 & 40809 & $\pm$ 144 & \\
& 3 & 18.41 & 40764 & $\pm$ 62 & \\
& 4 & 18.59 & 54726 & $\pm$ 131 & \\
& 5 & 19.18 & 55821 & $\pm$ 271 & \\
& 6 & 16.15 & 55316 & $\pm$ 123 & \\
& 7 & 16.59 & 40492 & $\pm$ 63 & R \\
& 8 & 18.63 & 41140 & $\pm$ 68 & \\
& 9 & 20.11 & - & - & \\
& 10 & 18.04 & - & - & \\
& 11 & 18.16 & - & - & $\ast$ \\
& 12 & 18.43 & 37199 & $\pm$ 151 & \\
& 13 & 17.60 & 40377 & $\pm$ 54 & \\
& 14 & 20.33 & - & - & \\
\hline\noalign{\smallskip}
{\bf 409N03} & 1 & 18.15 & 64532 & $\pm$ 127 & \\
& 2 & 17.44 & 47583 & $\pm$ 429 & \\
& 3 & 16.38 & 41309 & $\pm$ 50 & R \\
& 4 & 20.01 & - & - & \\
& 5 & 20.15 & 46345 & $\pm$ 145 & \\
& 6 & 17.65 & 46765 & $\pm$ 61 & \\
& 7 & 15.89 & 47150 & $\pm$ 55 & \\
& 8 & 18.30 & 46879 & $\pm$ 57 & \\
& 9 & 20.31 & - & - & \\
& 10 & 19.46 & 47460 & $\pm$ 247 & \\
& 11 & 19.83 & 46630 & $\pm$ 457 & \\
& 12 & 19.62 & - & - & \\
\hline\noalign{\smallskip}
{\bf 409N15} & 1 & 20.06 & 97958 & $\pm$ 75 & \\
& 2 & 17.44 & 45837 & $\pm$ 39 & \\
& 3 & 18.74 & 45687 & $\pm$ 42 & \\
& 4 & 18.54 & 45682 & $\pm$ 46 & \\
& 5 & 16.62 & 45280 & $\pm$ 31 & R \\
& 6 & 20.43 & 85363 & $\pm$ 98 & E \\
& 7 & 19.00 & 45343 & $\pm$ 58 & \\
& 8 & 20.53 & 38136 & $\pm$ 79 & \\
& 9 & 20.67 & 43891 & $\pm$ 100 & E \\
& 10 & 19.98 & 78625 & $\pm$ 65 & \\
& 11 & 19.59 & 79383 & $\pm$ 165 & \\
\hline\noalign{\smallskip}
{\bf 409N44} & 1 & 19.80 & 98238 & $\pm$ 172 & \\
& 2 & 20.02 & 69000 & $\pm$ 200 & \\
& 3 & 19.91 & 41416 & $\pm$ 158 & \\
& 4 & 17.62 & 40147 & $\pm$ 61 & R \\
& 5 & 19.69 & 41300 & $\pm$ 300 & \\
& 6 & 18.07 & 58828 & $\pm$ 73 & \\
& 7 & 17.71 & 40332 & $\pm$ 74 & \\
& 8 & 18.18 & 39350 & $\pm$ 42 & \\
\hline\noalign{\smallskip}
\end{tabular}
\end{flushleft}
\end{table}
\begin{table}
\begin{flushleft}
\addtocounter{table}{-1}
\caption{(Continued).}
\begin{tabular}{l@{\extracolsep{3mm}}l@{\extracolsep{6mm}}r@{\extracolsep{6mm}}c@{\extracolsep{6mm}}c@{\extracolsep{6mm}}c}
\hline
\hline\noalign{\smallskip}
NAME & N & $m_\mathrm{r}$ & $v$ & $\sigma_\mathrm{v}$~~~ & Notes \\
& & & ${\rm (km~s^{-1})}$ & ${\rm (km~s^{-1})}$ & \\
\hline
\hline\noalign{\smallskip}
& 9 & 18.26 & 58805 & $\pm$ 195 & \\
\hline\noalign{\smallskip}
{\bf 412N23} & 1 & 20.40 & 29350 & $\pm$ 300 & \\
& 2 & 19.81 & 49474 & $\pm$ 110 & \\
& 3 & 17.69 & 77698 & $\pm$ 144 & R \\
& 4 & 17.13 & 48980 & $\pm$ 80 & \\
& 5 & 19.14 & 32075 & $\pm$ 200 & \\
& 6 & 16.88 & 30590 & $\pm$ 300 & \\
& 7 & 18.78 & 76139 & $\pm$ 98 & \\
& 8 & 20.83 & - & - & \\
& 9 & 19.84 & 47521 & $\pm$ 78 & E \\
& 10 & 19.07 & 29934 & $\pm$ 57 & \\
& 11 & 19.52 & 60417 & $\pm$ 300 & \\
\hline\noalign{\smallskip}
{\bf 475N50} & 1 & 18.99 & 61683 & $\pm$ 84 & \\
& 2 & 19.86 & 64065 & $\pm$ 300 & \\
& 3 & 18.24 & 62720 & $\pm$ 300 & \\
& 4 & 18.59 & 63050 & $\pm$ 300 & \\
& 5 & 17.98 & 61318 & $\pm$ 94 & \\
& 6 & 17.34 & 63758 & $\pm$ 130 & \\
& 7 & 17.98 & 64280 & $\pm$ 173 & \\
& 8 & 17.49 & 63160 & $\pm$ 108 & R \\
& 9 & 20.79 & - & - & \\
& 10 & 19.62 & 63421 & $\pm$ 124 & \\
& 11 & 19.47 & - & - & $\ast$ \\
& 12 & 19.33 & 64246 & $\pm$ 118 & \\
\hline
\end{tabular}
\end{flushleft}
\end{table}

\section{Results}\label{sec:results}

On the basis of the measured galaxy velocities for each candidate cluster, we 
verified in which cases the spectroscopic data confirm the presence of a
cluster associated to a NVSS radiogalaxy.
As can be noticed from Table~\ref{tab:allvel}, for the candidate 349N02 the few
available spectroscopic data are not useful for a statistical analysis aimed to
assess the presence of a cluster around the radio source.

Among the $11$ fields for which we have sufficient data, in two cases (409N03 
and 412N23) the radiogalaxy velocity is significantly different from all the 
other measured values and we conclude that the radiogalaxy is not associated to
a cluster. In both cases the data suggest the presence of a group or cluster, 
but at a redshift different from that of the radiogalaxy.

For the $9$ remaining candidates, we confirm the presence of a cluster around 
the radiogalaxy: this corresponds to a positive detection rate of $82\%$. For 
these $9$ clusters we determined the mean velocity and velocity dispersion by 
means of the ROSTAT package (Robust Statistics, Beers et al. \cite{Beers}), 
which allows robust estimates of central location and scale in data samples 
affected by the presence of ``outliers''.
When dealing with small data sets  ($n = 5 - 10$) as in our case, the best 
estimators are the biweight $C_\mathrm{BI}$ (Tukey \cite{Tukey}) for the 
central location and the classical standard deviation $S_\mathrm{G}$ for the 
scale (Beers et al. \cite{Beers}).
The $C_\mathrm{BI}$ estimator is evaluated iteratively, by minimizing a 
function of the deviations of each observation from the estimate of the central
location. It thus requires an additional estimate of this last quantity, which 
is generally given as the absolute value of the median of the differences 
between the data and the sample median.

The uncertainties associated to central location and scale have been estimated
by the bootstrap method. This technique consists in the generation of a large 
number of samples, not independent from the original data set, and in the 
evaluation of the statistical parameters for each of these ``bootstrapped'' 
samples.

\begin{figure*}
\resizebox{\hsize}{!}{\includegraphics{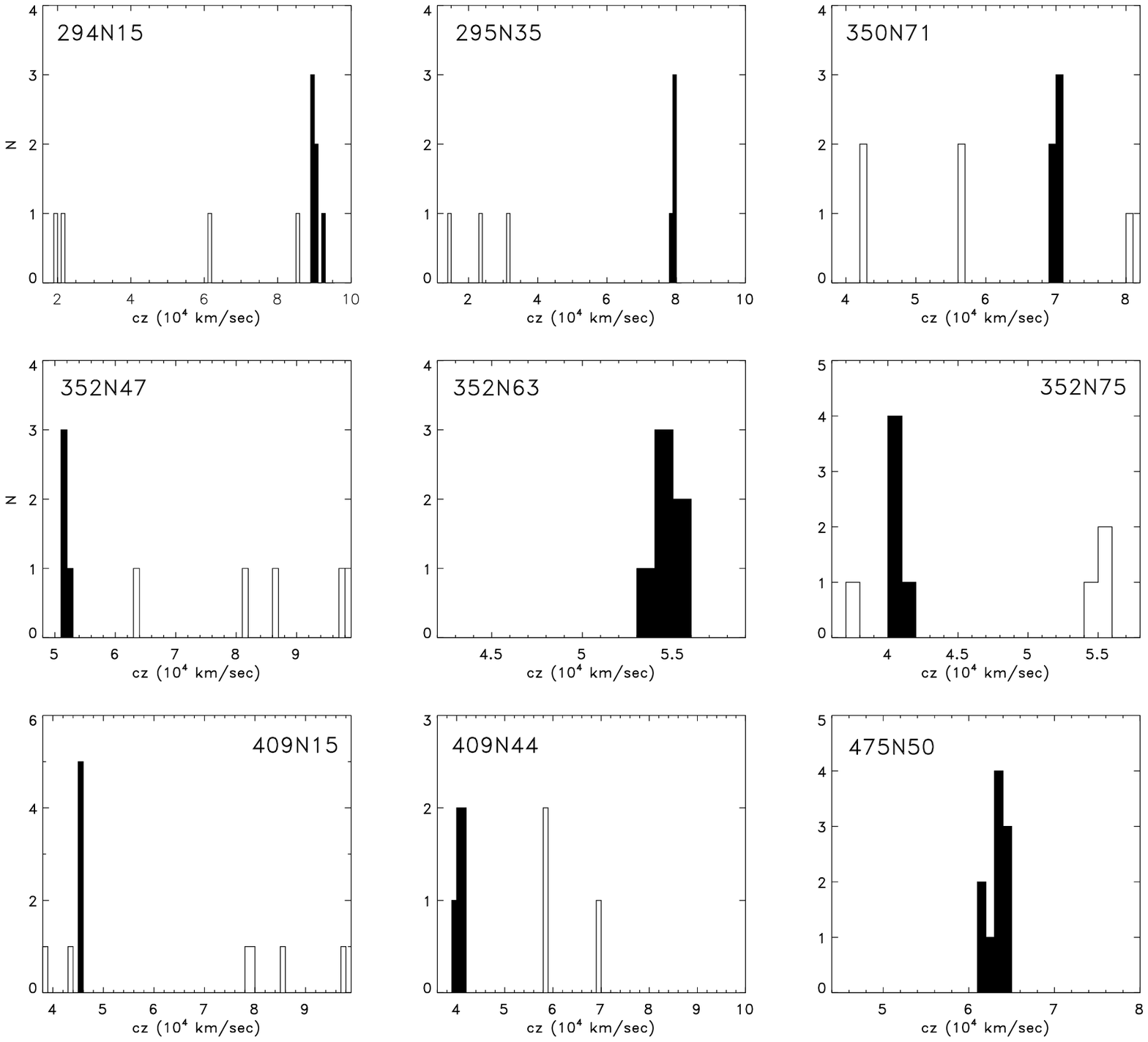}}
\caption{Measured velocities distributions for the $9$ spectroscopically
confirmed clusters. In black are shown the data sets used for the evaluation
of cluster redshift and velocity dispersion (see Table~\ref{tab:myclus}).}
\label{fig4}
\end{figure*}

In Fig.~\ref{fig4} we show the distributions of measured velocities for the
$9$ cluster candidates involved in this statistical analysis: the shadowed 
regions represent the data sets used as input for the ROSTAT package.

The results of the statistical analysis are shown in Table~\ref{tab:myclus}: 
mean cluster velocities vary from $40514~{\rm km~s^{-1}}$ to 
$90122~{\rm km~s^{-1}}$, corresponding to the redshift range 
$0.13 \le z \le 0.3$.

Despite the small number of available redshifts for each cluster, which
reflects into rather large errors for both the central location and velocity 
dispersion, an interesting result arises from the velocity dispersions: they 
range from $210~{\rm km~s^{-1}}$ to $906~{\rm km~s^{-1}}$, that is from values 
typical of poor clusters or groups of galaxies to those typical of moderately 
rich clusters.

Following the criteria in Abell 
(\cite{Abell58}), we used the EDSGC catalog to get an estimate of the cluster 
richness for the $9$ confirmed clusters: the background--subtracted galaxy 
counts in the magnitude range $m_3 \div m_3 +2$ within an Abell radius from the
cluster centre range from a minimum of $6$ to a maximum of $23$.
These galaxy counts are similar to those found for many of the ACO poor 
clusters (Abell et al. \cite{Abell}), and suggest that our radio--optically 
selected clusters are poorer than Abell richness class $0$.
We stress however that these richness estimates must be viewed with caution:
first, the values of $m_3 +2$ often fall near or below $b_J=20.5$, where the 
EDSGC completeness drops significantly, thus seriously biasing the galaxy 
counts. Second, at our typical $m_3$ the number density of galaxies in the 
EDSGC is high, about $50$ galaxies per square degree, thus the probability of 
selecting as the third member of the cluster a galaxy which is actually a 
background or foreground object seen in projection is not negligible, and 
this again can alter the richness estimate.

\begin{figure}
\resizebox{\hsize}{!}{\includegraphics{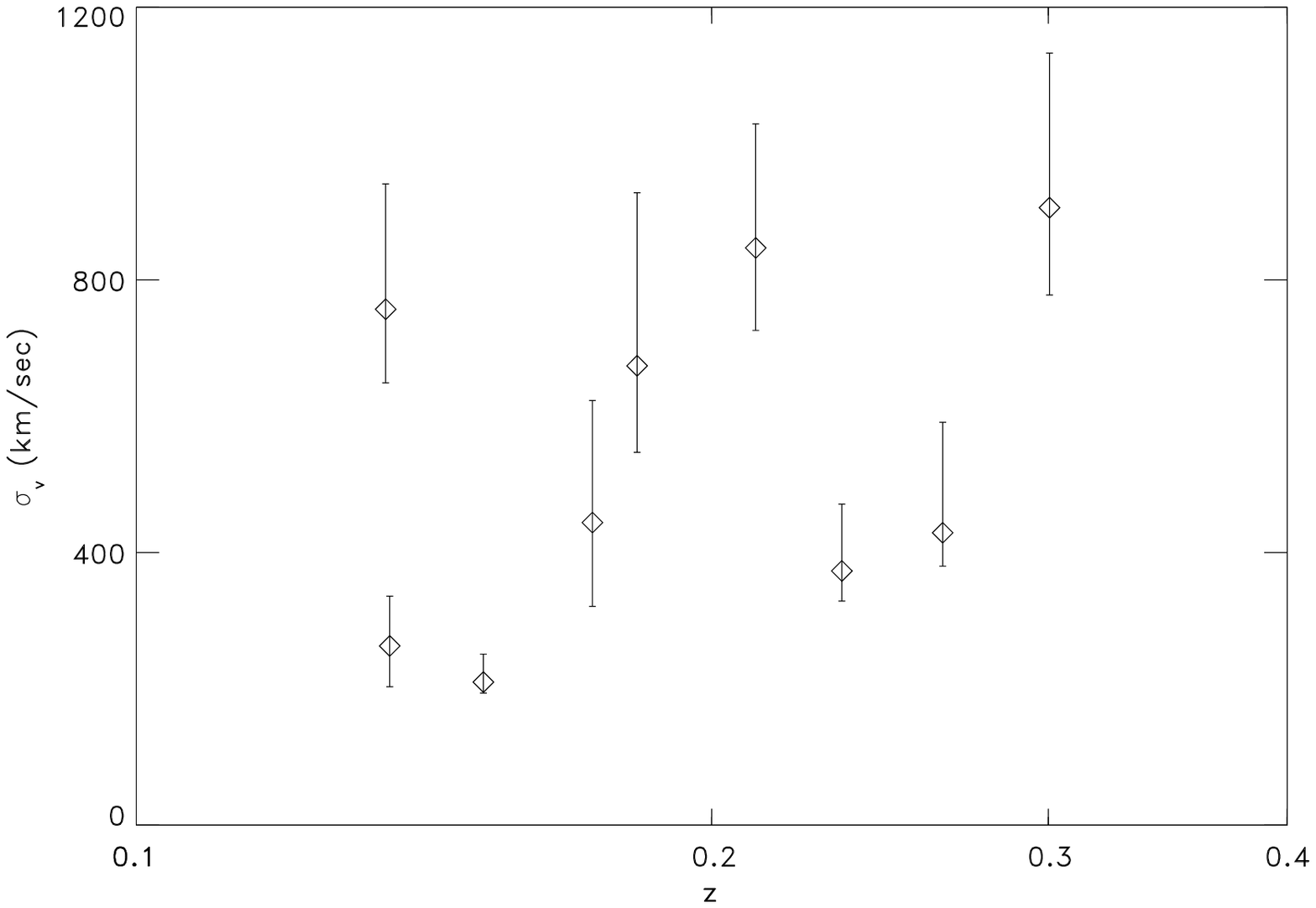}}
\caption{Velocity dispersion versus redshift for the $9$ spectroscopically
confirmed clusters. Error bars are $1 \sigma$.
The lack of correlation between redshift and velocity dispersion suggests that 
structures with different richness are well represented at any distance in the 
redshift range covered by our candidate cluster sample.}
\label{fig5}
\end{figure}

As shown in Fig.~\ref{fig5}, there is no evident correlation between measured
velocity dispersion and cluster redshift. The use of radio emission properties
of galaxies seems thus a very efficient method to select new candidate clusters
samples in a wide range of richness at any redshift.

\begin{table*}
\caption{Mean velocity and velocity dispersion obtained from the package ROSTAT
for the $9$ spectroscopically confirmed clusters. Values for $<v>$ are given by
the biweight estimator of central location, while for $\sigma_\mathrm{v}$ the 
classical standard deviation is given: these are the most robust and efficient 
statistical estimators in the case of small data sets. The number of measured 
velocities used to obtain these results is given in column 4.
\label{tab:myclus}}
\begin{flushleft}
\begin{tabular}{cccccrccccccc} 
\hline
\hline\noalign{\smallskip}
CLUSTER & & Right Ascension (B1950) & & Declination (B1950) & & n & & & $<v> ~{\rm (km~s^{-1})}$ & & & $\sigma_\mathrm{v} ~{\rm (km~s^{-1})}$ \\
\noalign{\smallskip}\hline
\hline\noalign{\smallskip}
294N15 & & 00 23 41.0 & & -39 37 15.0 & & 6 & & & 90122~$^{+519}_{-589}$ & & & 906~$^{+227}_{-128}$  \\
\noalign{\smallskip}\hline\noalign{\smallskip}
295N35 & & 01 03 11.0 & & -38 47 15.0 & & 4 & & & 79241~$^{+130}_{-270}$ & & & 429~$^{+162}_{-49}$  \\
\noalign{\smallskip}\hline\noalign{\smallskip}
350N71 & & 00 35 04.0 & & -34 49 15.0 & & 5 & & & 70180~$^{+293}_{-138}$ & & & 373~$^{~+98}_{~-44}$  \\
\noalign{\smallskip}\hline\noalign{\smallskip}
352N47 & & 01 14 13.0 & & -36 44 45.0 & & 4 & & & 51969~$^{+14}_{-309}$ & & & 444~$^{+179}_{-123}$  \\
\noalign{\smallskip}\hline\noalign{\smallskip}
352N63 & & 01 19 50.0 & & -33 45 15.0 & & 6 & & & 54844~$^{+497}_{-144}$ & & & 674~$^{+254}_{-127}$  \\
\noalign{\smallskip}\hline\noalign{\smallskip}
352N75 & & 01 21 11.0 & & -33 17 15.0 & & 5 & & & 40712~$^{~+66}_{-227}$ & & & 263~$^{~+73}_{~-60}$  \\
\noalign{\smallskip}\hline\noalign{\smallskip}
409N15 & & 00 02 36.0 & & -28 20 15.0 & & 5 & & & 45573~$^{+96}_{-254}$ & & & 210~$^{~+41}_{~-16}$  \\
\noalign{\smallskip}\hline\noalign{\smallskip}
409N44 & & 23 51 13.0 & & -31 34 15.0 & & 5 & & & 40514~$^{+761}_{-269}$ & & & 757~$^{+184}_{-108}$  \\
\noalign{\smallskip}\hline\noalign{\smallskip}
475N50 & & 01 15 13.0 & & -24 09 45.0 & & 10 & & & 63266~$^{+256}_{-414}$ & & & 847~$^{+182}_{-121}$  \\
\noalign{\smallskip}\hline\noalign{\smallskip}
\end{tabular}
\end{flushleft}
\end{table*}

If confirmed by future spectroscopic follow-up, this result could be of great 
interest as our sample would offer the possibility to investigate differences 
in cluster dynamical properties in a homogeneously selected sample of clusters
which spans a wide range in richness, and to improve our knowledge of their 
number counts, as well as to study the radio emission properties of galaxies 
residing in different environments.

\section{Conclusions}\label{sec:concl}

To study the status and the evolution of clusters of galaxies at intermediate
redshifts we built a sample of candidate clusters using radiogalaxies in the 
NRAO VLA Sky Survey as tracers of dense environments.

From the NVSS maps we extracted a catalogue of radio sources over an area of 
$\approx 550$ square degrees, and made optical identifications with galaxies 
brighter than $b_\mathrm{J} = 20.0$ in the EDSGC Catalogue, resulting in a 
sample of $1288$ radiogalaxies (Zanichelli et al. \cite{Zanichelli}, Paper I).

In this paper we have presented the detection technique we applied to select
candidate groups and clusters associated to NVSS radio sources. The method is
based on the search of excesses in optical surface galaxy density nearby NVSS 
radiogalaxies. 
To keep low the probability of spurious radio-optical identifications, as well 
as to preferentially select clusters at redshifts $z \simgt 0.1$, we restricted
the cluster search to the $661$ radiogalaxies having radio-optical distance 
$\le 7\arcsec$ and magnitude  $b_\mathrm{J}\ge 17.5$.

The search of regions having high optical galaxy density has been made using 
the EDSGC galaxy catalogue, building matrices of galaxy counts down to
magnitude $b_\mathrm{J} = 20.5$.
This choice allows to find density excesses surrounding the faintest 
radiogalaxies (identified down to $b_\mathrm{J} = 20$) without introducing 
significant incompleteness effects in the optical data.
Smoothing of galaxy counts has been done using a gaussian filter with 
FWHM$ = 2\arcmin$. The mode and standard deviation of smoothed galaxy counts 
have been used to define a detection threshold for the surface density 
excesses: we selected as cluster candidates those density excesses whose
centroid is within $4\arcmin$ from a radiogalaxy.
This search radius for candidate clusters associated to NVSS radio sources
corresponds to an Abell radius of a cluster at $z \sim 0.45$.

By applying this cluster detection strategy to $661$ radiogalaxies over 
$\approx 550$~sq.~degrees at the South Galactic Pole, we obtained a sample of 
$171$ cluster candidates. The estimated contamination level is about $28\%$.

Out of these $171$ candidates, $76$ correspond to already known clusters, while
$95$ cluster candidates in our list do not have any known counterpart in the 
literature and have been the subject of subsequent spectroscopic follow-up. The
full sample of radio-optically selected cluster candidates will be presented in
a following paper.
Multi Object Spectroscopy aimed to confirm the detection of clusters has been
successfully acquired at the 3.6 m ESO telescope for a subset of $12$ 
candidates.
In $2$ cases the radiogalaxy does not lie at the same redshift as any other 
observed target, while $9$ candidates have been confirmed as clusters of 
galaxies in the redshift range $0.13 \le z \le 0.3$, thus confirming that this 
joint radio-optical cluster selection technique can be used as a powerful tool 
for the detection of cluster candidates at intermediate redshifts.
For one additional candidate, the very low number of measured redshifts does 
not allow any conclusion on the presence of a cluster surrounding the 
radiogalaxy, and further observations are needed.
Velocity dispersions of the $9$ spectroscopically confirmed clusters vary from 
values typical of moderately rich clusters to those typical of groups or poor 
clusters, thus strengthening the assumption that this technique is equally 
efficient in selecting structures over a wide range of richness at different 
redshifts.
If confirmed by future spectroscopic follow up, this last result could be of
great interest as this technique would offer the possibility to study the
properties of different environments, such as groups or rich clusters, in a
homogeneously selected cluster sample.

\begin{acknowledgements}
The authors acknowledge Marco Mignoli for his valuable help during the 
observative run.
\end{acknowledgements}


\begin{thebibliography}{}

\bibitem[1958]{Abell58}
Abell G.O, 1958, ApJS 3, 211

\bibitem[1989]{Abell}
Abell G.O, Corwin H.G., Olowin R.P., 1989, ApJS 70, 1

\bibitem[1993]{Allington--Smith}
Allington--Smith J.R., Ellis R.S., Zirbel E.L., Oemler A., 1993, ApJ 404, 521

\bibitem[1990]{Beers}
Beers T.C., Flynn K., Gebhardt K., 1990, AJ 100, 32

\bibitem[1994]{Burns}
Burns J.O., Rhee G., Owen F.N., Pinkney J., 1994, ApJ 423, 94

\bibitem[1984]{Butcher}
Butcher H.R., Oemler A., 1984, ApJ 285,426

\bibitem[1992]{Collins92}
Collins C.A., Nichol R.C., Lumsden S.L., 1992, MNRAS 254, 295

\bibitem[1995]{Collins95}
Collins C.A., Guzzo L., Nichol R.C., Lumsden S.L., 1995, MNRAS 274, 1071

\bibitem[1998]{Condon}
Condon J.J., Cotton W.D., Greisen E.W., et al., 1998, AJ 115, 1693

\bibitem[1994]{Dalton94}
Dalton G.B., Efstathiou G., Maddox S.J., Sutherland W.I., 1994, MNRAS 269, 151

\bibitem[1997]{Dalton97}
Dalton G.B., Maddox S.J., Sutherland W.I., Efstathiou G., 1997, MNRAS 289, 263

\bibitem[1982]{Feigelson}
Feigelson E.D., Maccacaro T., Zamorani G., 1982, ApJ 255, 392

\bibitem[1994]{Frei}
Frei Z., Gunn J.E., 1994, AJ 108, 147

\bibitem[1990]{Gioia}
Gioia I.M., Henry J.P., Maccacaro T., et al., 1990, ApJ 356, L35

\bibitem[1977]{Grueff}
Grueff G., Vigotti M., 1977, A\&A 54, 475

\bibitem[1992]{Henry}
Henry J.P., Gioia I.M., Maccacaro T., Morris S.L., Stocke J.T., 1992,
ApJ 386, 408

\bibitem[1991]{Hill}
Hill G.J., Lilly S.J., 1991, ApJ 367, 1

\bibitem[1996]{Katgert96}
Katgert P., Mazure A., Perea J., et al., 1996, A\&A 310, 8

\bibitem[1998]{Katgert98}
Katgert P., Mazure A., den Hartog R., et al., 1998, A\&AS 129, 399

\bibitem[1996]{Ledlow}
Ledlow M.J., Owen F.N., 1996, AJ 112, 9

\bibitem[1992]{Lumsden}
Lumsden S.L., Nichol R.C., Collins C.A., Guzzo L., 1992, MNRAS 258, 1

\bibitem[2000]{Nichol}
Nichol R.C., Collins C.A., Lumsden S.L., 2000, submitted to ApJS

\bibitem[1996]{Postman}
Postman M., Lubin L.M., Gunn J.E., et al., 1996, AJ 111, 615

\bibitem[1988]{Prestage}
Prestage R.M., Peacock J.A., 1988, MNRAS 230, 131

\bibitem[1999]{Ramella}
Ramella M., Zamorani G., Zucca E., et al., 1999, A\&A 342, 1

\bibitem[1998]{Rosati}
Rosati P., della Ceca R., Norman C., Giacconi R., 1998, ApJ 492, L21

\bibitem[1999]{Scodeggio}
Scodeggio M., Olsen L.F., da Costa L., et al., 1999, A\&A 137, 83

\bibitem[1996]{Shectman}
Shectman S.A., Landy S.D., Oemler A., et al., 1996, ApJ 470, 172

\bibitem[1991]{Stocke}
Stocke J.T., Morris S.L., Gioia I.M., et al., 1991, ApJS 76, 813

\bibitem[1999]{Struble}
Struble M.F., Rood H.J., 1999, ApJS 125, 35

\bibitem[1958]{Tukey}
Tukey J.W., 1958, Ann. Math. Stat. 29, 614

\bibitem[1997]{Vettolani97}
Vettolani G., Zucca E., Zamorani G. et al., 1997, A\&A 325, 954

\bibitem[1998]{Vettolani98}
Vettolani G., Zucca E., Merighi R., et al., 1998, A\&AS 130, 323

\bibitem[2001]{Zanichelli}
Zanichelli A., Vigotti M., Scaramella R., et al., 2001, 
A\&A in press, Paper I

\bibitem[1989]{Zhao}
Zhao J.H., Burns J.O., Owen F.N., 1989, AJ 98, 64

\bibitem[1996]{Zirbel96}
Zirbel E.L., 1996, ApJ 473, 713

\bibitem[1997]{Zirbel97}
Zirbel E.L, 1997, ApJ 476, 489

\end{thebibliography}
\end{document}